\documentclass{sig-alternate}

\usepackage[center]{subfigure}
\usepackage{graphicx}
\usepackage{hyperref}
\usepackage{rotating}
\usepackage{supertabular}
\usepackage{multirow}
\usepackage{moresize}
\usepackage{makeidx}
\usepackage{color}
\usepackage{etoolbox}

\usepackage{blindtext}
\makeatletter
\patchcmd{\maketitle}{\@copyrightspace}{}{}{}
\makeatother

\begin{document}
\title{Frapp\'e: Understanding the Usage and Perception of Mobile App Recommendations In-The-Wild}

\numberofauthors{4}
\author{
\alignauthor Linas Baltrunas\\
\affaddr{Telefonica Research}\\
\email{linas@tid.es}
\alignauthor Karen Church\\
\affaddr{Yahoo Labs}\\
\email{kchurch@yahoo-inc.com}
\and 
\alignauthor Alexandros Karatzoglou\\
\affaddr{Telefonica Research}\\
\email{alexk@tid.es}
\and
\alignauthor Nuria Oliver\\
\affaddr{Telefonica Research}\\
\email{nuriao@tid.es}
}
\maketitle

\begin{abstract}
This paper describes a real world deployment of a context-aware mobile app recommender system (RS) called Frapp\'e. Utilizing a hybrid-approach, we conducted a large-scale app market deployment with 1000 Android users combined with a small-scale local user study involving 33 users. The resulting usage logs and subjective feedback enabled us to gather key insights into (1) context-dependent app usage and (2) the perceptions and experiences of end-users while interacting with context-aware mobile app recommendations. While Frapp\'e performs very well based on usage-centric evaluation metrics insights from the small-scale study reveal some negative user experiences. Our results point to a number of actionable lessons learned specifically related to designing, deploying and evaluating mobile context-aware RS in-the-wild with real users. 
\end{abstract}


\begin{figure*}[t]
  \begin{center}
    \subfigure[Main Screen with 3 most relevant recommendations.]{\label{fig:ss2}
      \includegraphics[width=0.18\linewidth]{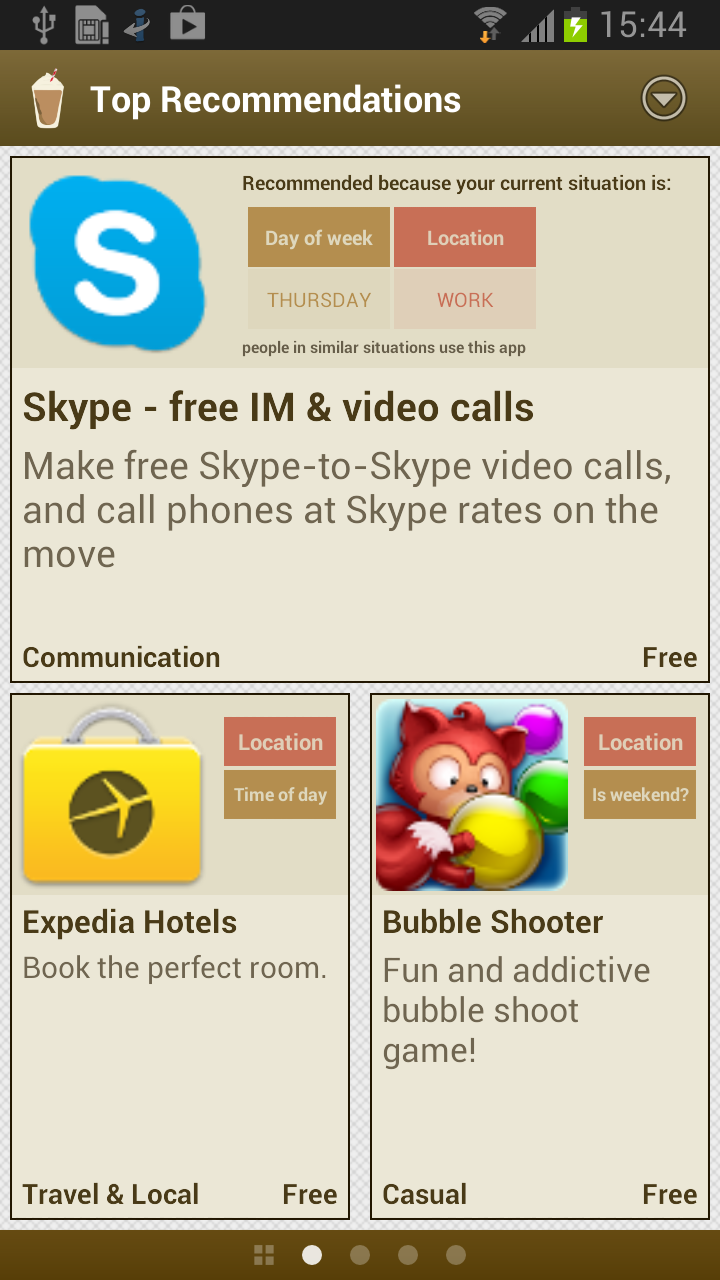}}
    \subfigure[Recommendation screen and options menu.]{\label{fig:ss6}
      \includegraphics[width=0.18\linewidth]{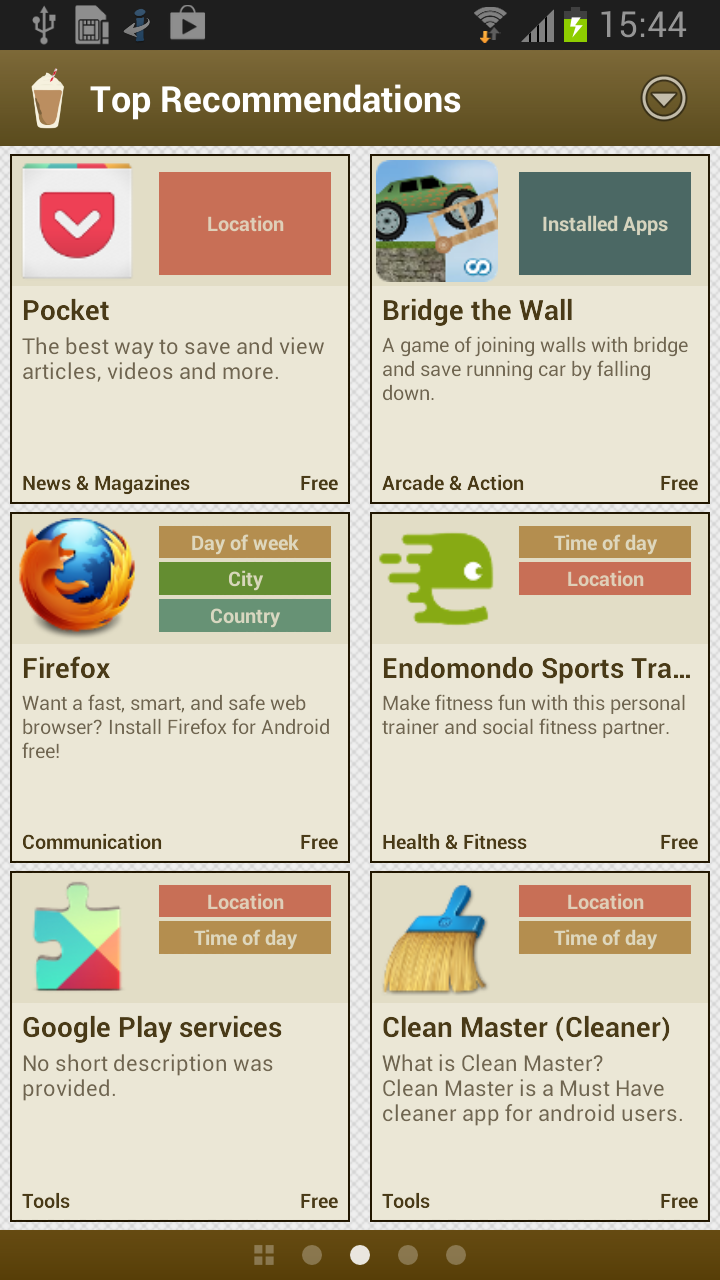}}
    \subfigure[Details screen for an app.]{\label{fig:ss7}
      \includegraphics[width=0.18\linewidth]{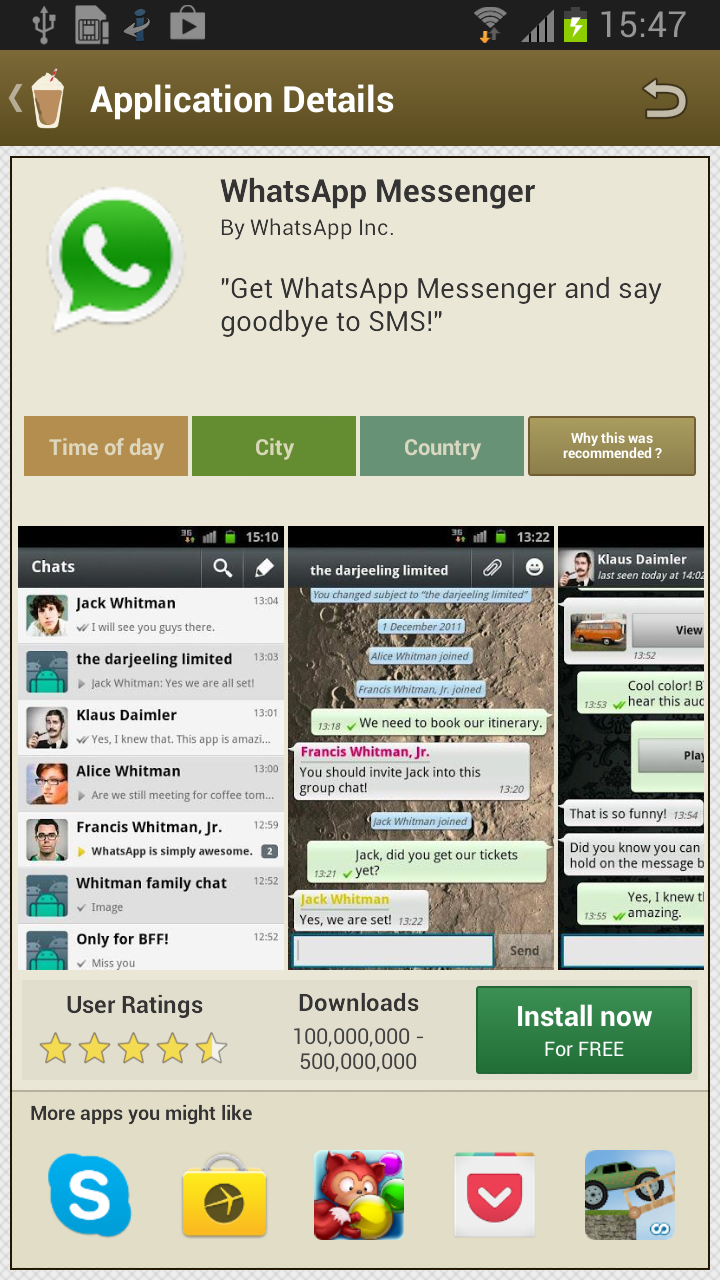}}
    \subfigure[Recommendation explanation example.]{\label{fig:ss8}
      \includegraphics[width=0.18\linewidth]{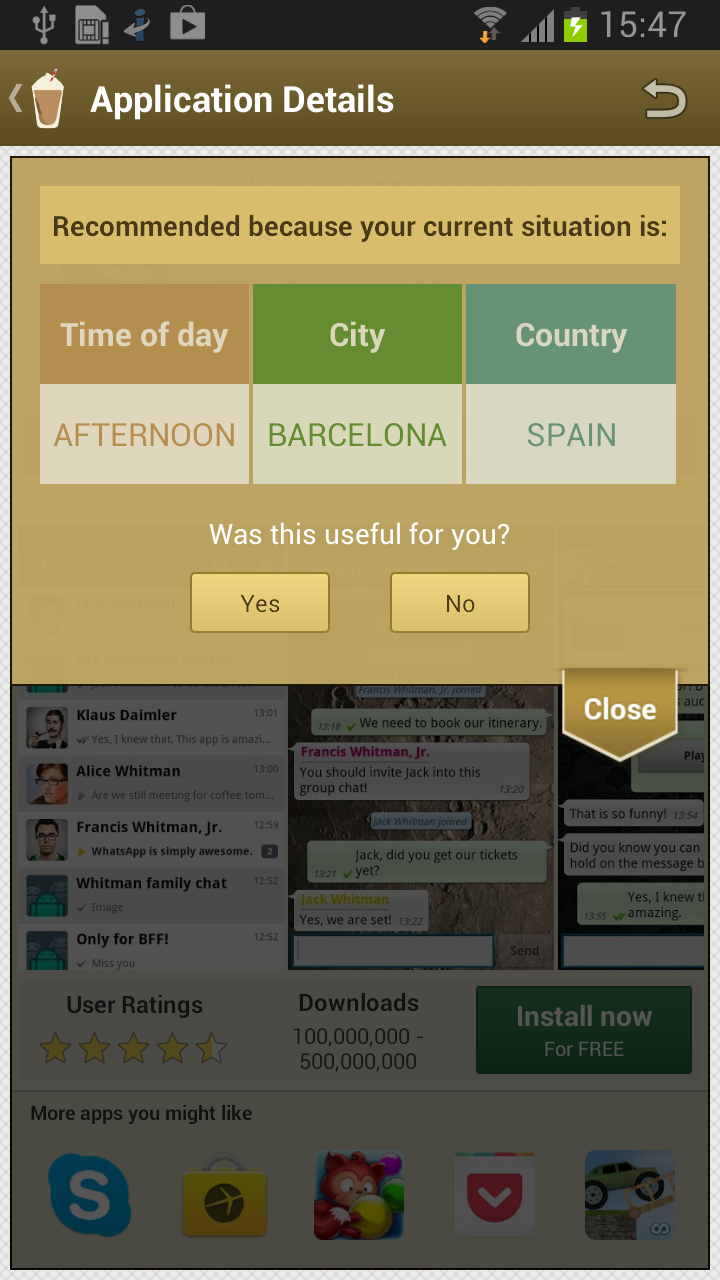}}
    \subfigure[Most relevant categories.]{\label{fig:ss4}
      \includegraphics[width=0.18\linewidth]{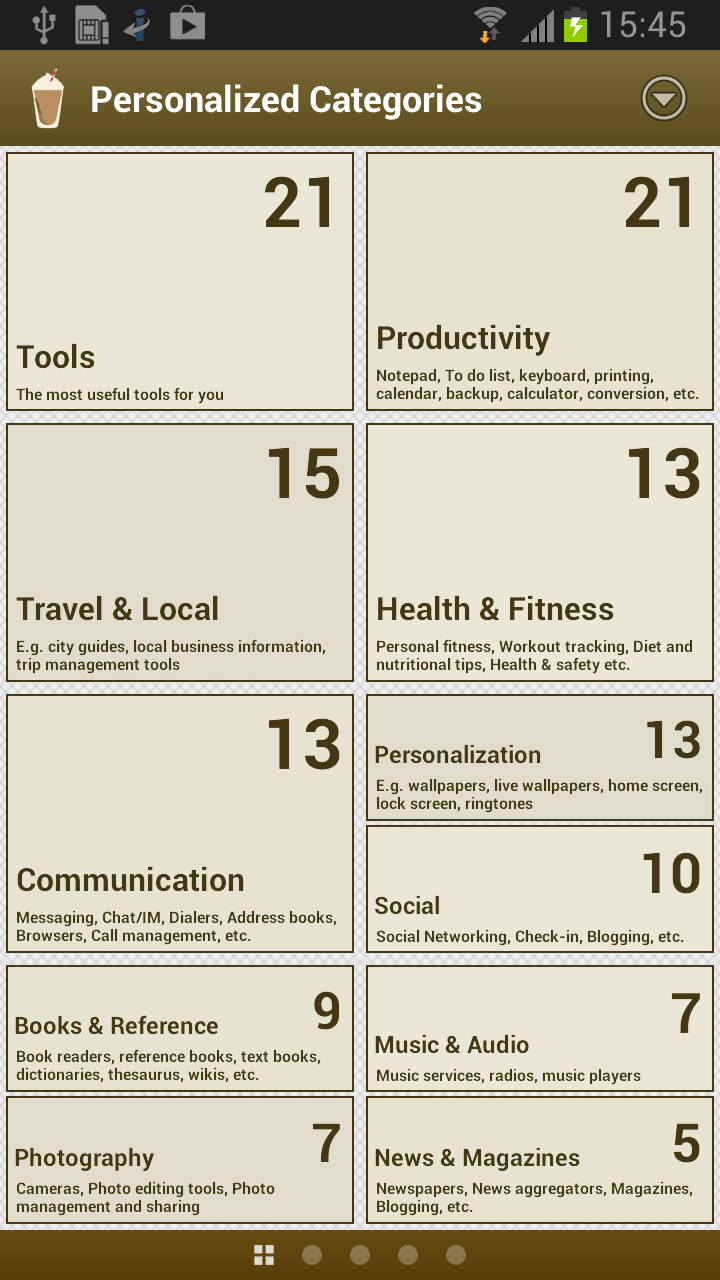}}
   \end{center}
  \caption{Screenshots of Frapp\'e.}
  \label{fig:ss}
\end{figure*}

\section{Introduction}

The mobile world continues to grow at a phenomenal rate. Once used as mere communications devices, mobile phones have evolved into sophisticated personal computers enabling access to a wealth of information and services, anytime, and anywhere. These smartphones now include sensors that enable the collection and analysis of contextual information (\emph{e.g.} GPS, accelerometer) from which we can draw insights about the intentions, activities and locations of mobile users. 
The world of mobile apps has also grown exponentially. Both Apple\footnote{See http://bit.ly/1cKxavd} and Google\footnote{See http://bit.ly/1jlVJQt} have recently reported reaching one million apps in their app stores. Users' demand for mobile apps is also steadily increasing, with downloads from mobile app stores expected to reach 98 billion by 2015\footnote{See: http://bit.ly/13UgmLZ}.

This increasing volume of available mobile apps has resulted in significant app overload and app discovery challenges for mobile users. 
As a result, several app recommendation and aggregation services have emerged. Some of these services support the discovery of relevant apps through either ratings or recommendations based on user profiles, types of installed apps, and in some cases the usage of those apps, \emph{e.g.} \emph{Appolicious, AppsFire, Zwapp, Appsaurus}, and \emph{AppAware} \cite{Girardello:2010}. Recent work has leveraged unique mobile contexts, such as location to provide \emph{context-aware recommendations} or CARS \cite{Adomavicius10-handbook, Baltrunas11-journal, Karatzoglou12-appwall}. These mobile contexts have been shown to have a significant impact on the needs, behaviors and app usage patterns of mobile users \cite{Church-2009,Bohmer11-angry}. Hence, it seems likely that utilizing such contextual data in mobile app recommendation  would lead to enhanced end-user experiences.  

Context-aware recommendation algorithms have been shown to outperform other state-of-the-art recommendation approaches. However, the vast majority of these evaluations were conducted off-line, with a core focus on \emph{performance} and \emph{effectiveness} from an algorithmic perspective. To date, little is understood about (1) how useful these context-aware algorithms are in a real-world, in-situ scenario, nor the (2) the subjective perceptions and experiences of end-users with the recommendations provided. 
 
In this paper we describe a real-world deployment of a mobile app recommender called Frapp\'e which provides context-aware mobile app recommendations by means of a Tensor Factorization approach \cite{Karatzoglou12-appwall}. Frapp\'e's recommendations leverage both implicit usage data and contextual factors to provide suggestions of relevant apps to each user.
The core contributions of the this work can be summarized as follows:
\begin{itemize} 
\item We released anonymised context-aware app usage data set, which can be found at \url{http://baltrunas.info/research-menu/frappe}
\item A characterization of context dependent app usage, based on the data gathered via our large-scale app market deployment of Frapp\'e in Google Play with 1000 Android users; 
\item Key results from our hybrid \emph{in-situ} study of Frapp\'e highlighting key insights into user experiences; 
\item A set of actionable lessons learned related to how to effectively design, deploy and evaluate context-aware mobile RS in-the-wild with real users.
\end{itemize}
\section{Related Work}

\textbf{Recommenders for Mobile Apps: }
Earlier work on mobile context-aware recommender systems (CARS) \cite{Baltrunas11-journal} has shown that
contextual factors, \emph{e.g.} time, location, activity, weather, emotional state and the user's social
network, heavily 
influence the recommendation needs of
people \cite{Adomavicius10-handbook,Bohmer:2013mobil}.  
Most of this research has focused on the performance and effectiveness of the recommender from an algorithmic and performance perspective, highlighting positive off-line evaluation results, e.g. ~\cite{Adomavicius10-handbook, Karatzoglou12-appwall}. 

Given the rise in popularity of mobile app markets like Google Play and Apple's App Store, coupled with the increasing volume of available mobile apps, researchers in the CARS space have begun to focus explicitly on the \emph{mobile app recommendation} domain.  For example, Woerndl et al. \cite{Woerndl:2007} describes a hybrid RS that can recommend mobile apps to users based on what other users have installed in a similar context, in this case location. 
Davison \& Moritz \cite{Davidsson:2011} present \emph{Applause} which provides context-aware recommendations utilizing location as the key form of context and provides mechanisms for solving the \emph{new user problem}.
In \cite{Jannach:2009}, Jannach \& Hegelich focus on recommendation of games applications within a mobile Internet portal and show that game buying behaviors increase among users who receive personalized recommendations vs. non-personalized recommendations. 
\emph{AppJoy} \cite{Yan-2011} supports personalized mobile app recommendations by combining item-based collaborating filtering with data on how the user actually uses his/her installed apps. Offline evaluation results using a dataset of 4600 Android users showed that users interacted longer with the recommended apps.
 \emph{AppAware} \cite{Girardello:2010} recommends new mobile applications by making use of context in the form of location. 
It provides the user with real-time information of application installs, uninstalls and updates so that the user is made more aware of what applications other users are interacting with in his/her proximity. 
\emph{Appazaar} \cite{Bohmer11-angry} is a prototype recommender system for mobile apps which utilizes the user's current and 
historical location information as well as app usage to recommend apps. Data gathered via a Google play deployment of Appazaar was used to provide in-depth insights 
about mobile app usage by smartphone users \cite{Bohmer11-angry}. Most recently Bohmer et al. \cite{DBLP:conf/iui/BohmerGK13} propose a \emph{usage-centric} framework to evaluate app recommenders by utilizing key events within an mobile app's life-cycle (\emph{e.g.} recommendation, install and long-term usage).

\textbf{User Studies of Recommender Systems (RS): }
In recent years, the community has begun to investigate RS effectiveness from a more user-centric perspective \cite{Konstan:2012}. McNee et al. \cite{McNee:2006} highlight that user satisfaction 
does not always correlate with high recommender 
accuracy and argue that RS evaluations should move beyond traditional accuracy metrics and look towards more user-centric factors. Pu et al. \cite{Pu:2011} outline a comprehensive framework to evaluate the \emph{perceived qualities of recommender systems} in a model called ResQue (Recommender 
systems' Quality of user experience). 
In \cite{Knijnenburg2012}, Knijnenburg et al. describes a framework for understanding the
user experience of RS, describing why and how certain aspects of a system lead to better user experiences while others do not.
In \cite{Cremonesi-2011-eval}, Cremonesi et al. report on an empirical study involving 210 users which explored the users  perceived quality of seven different RSs on the same dataset in an offline evaluation. 

While user-centric evaluations of RS have been conducted, the majority of these studies have taken place in lab-based settings, with few participants and no concrete mobile focus. To date, real-world deployments of mobile RS in-the-wild have been rather limited, with very few subjective insights from users.  The goal of this paper is to help bridge this gap by combining a large scale deployment of a mobile app recommender, with a smaller scale in-situ field study to gather interesting insights into the experiences and perceptions of context-aware app usage and recommendation \emph{in-the-wild}. To the best of our knowledge this work is the first of its kind in the mobile RS space because of the hybrid approach, its scope and scale.

\section{Frapp\'e}
\label{sec:frappe}

\begin{table}[htbp]
\scriptsize
\begin{tabular}{p{1.7cm}|p{6.2cm}}
Input Signal &  Values \\
\hline
Installed apps      	& All non-system installed applications on the Android phone. \\
Used apps           	& \# times application was used (in a specific context) \\
Skipped apps       	& Apps that were recommended but not viewed/installed by the user \\
Viewed apps         	& Apps that were recommended and installed by the user \\
\hline
Time of the day     & One of 4 possible values: Morning (6am to 12am), Afternoon (12am to  6pm) Evening (6pm to 12pm), Night (12pm to 6am)\\
Day of the week     & One of 7 possible values: Mon, Tues, Weds, Thurs, Fri, Sat or Sun \\
Period of the week  & One of 2 possible values: Weekend or Working day (\emph{i.e.} weekday) \\
Location            & One of 3 possible values: Home, Work, Other \\
City                & Boolean: True, if user is close (20km) to the center of a major city; False otherwise \\
Country             & Name of the country where the user is currently located \\
Weather             & One of 9 possible values: Sunny, Cloudy, Foggy, Windy, Drizzle, Rainy, Stormy, Sleet, Snowy \\
\hline
Battery Level       & One of 5 possible values: Full, High, Medium, Low, Empty \\
Energy Source       & One of 3 possible values: Battery, USB, AC \\
Connectivity        & One of 3 possible values: WiFi, Mobile, No \\
Screen State        & Boolean: True, on; False, off \\
\end{tabular}
\caption{Input signals used in Frapp\'e engine. \label{tab:inputs}}
\end{table}

Frapp\'e is a context-aware personalized recommender of mobile
apps. It runs on Android phones and recommends the most
relevant apps for the user based on his/her current situation (context) and usage patterns. Frapp\'e
automatically adapts to the user's needs by utilizing information regarding installed and used
apps in a variety of contexts and settings. The full list of currently supported
contextual input signals used by Frapp\'e is shown in Table~\ref{tab:inputs}. 
Frapp\'e's recommendations are provided by a novel CARS algorithm described in~\cite{Karatzoglou12-appwall}, which uses a Tensor Factorization approach. 
The model was designed to work with \emph{implicit feedback} data for mobile app
recommenders. In this case, the implicit data contains information about how often a user used an app, for how long and 
in which contexts. However, we do not have explicit user feedback, \emph{e.g.} their rating for a given app.
Therefore, we consider app usage as an indirect indication of user interest in the app.
Frapp\'e's algorithm was shown to achieve up to a 28\% improvement in performance (measured in Mean Average Precision) over another state-of-the-art
method presented in \cite{koren08-implicit} in a series of off-line experiments using a dataset generated by the Appazaar\footnote{See: http://appazaar.net/} 
mobile app recommender.

\textbf{Architecture and Logging: }
Frapp\'e uses a standard client-server architecture. The
server side computes the recommendation models, and provides the top-21
recommendations given a client request via the HTTPS protocol. 
The client side consists of an Android mobile app that (1) displays the recommendations and (2) runs a background service, 
which collects app usage data and other contextual information as shown in Table \ref{tab:inputs}.
This background service in the client, similar to \cite{Bohmer11-angry}, samples the current
state of the phone once per minute and gathers (1) the last known readings of various
sensors and (2) information on which applications are currently in use (in the foreground) by the
user. This data is sent in batches to the Frapp\'e server where it is cleaned
and enriched with additional information such as weather and 
location information abstracted at a higher level (\emph{i.e.}, home/work, country, city, etc).
This architecture provides
detailed logs on app usage with minimal impact on battery consumption ( $<3\%$ of daily battery consumption).

\textbf{The Frapp\'e Interface:}
Screenshots of Frapp\'e can be
seen in Figure~\ref{fig:ss}. The main screen of Frapp\'e (
Figure~\ref{fig:ss2}) shows the top 3 most relevant recommendations for the
current user in their current context. Users can swipe to the right and get more
recommendations (see Figure~\ref{fig:ss6}). If the user clicks on the tile of an
application, the app details screen appears 
(Figure~\ref{fig:ss7}). If the user decides to install an app (s)he
is redirected to the Google play website for that app. If the user
swipes to the far left, (s)he is presented with personalized
categories (see Figure~\ref{fig:ss4}), \emph{i.e.} the list of categories that are
most relevant to that particular user in their current context. Frapp\'e also logs all
user interactions and clicks inside the app. Therefore, we can trace back and observe which apps
the user viewed or installed and in what situations.

\textbf{Explanations of Recommendations: }
In addition to recommendations, Frapp\'e also provides an explanation of the contextual factors 
that lead to the recommendation of each app (See Figure~\ref{fig:ss8} for an example.).
The main purpose of these explanations is to communicate
the role of context in making the recommendations, \emph{i.e.}, to share with the user 
the fact that Frapp\'e takes their context into account and to explain which 
specific contextual factors were the most important when making that particular recommendation. 

Frapp\'e uses a chi-square based
heuristic to generate explanations. The explanation engine exploits app usage statistics for all users 
to generate plausible reasons for the recommendation. Here we assume independence of the contextual factors.
If an app is used significantly more often in a specific context than the average app, 
this contextual factor is used to explain the recommendation. We first compute the
Pearson's statistic:
\[
\chi^2 = \frac{(O_{ic} - E_{ic})^2}{E_{ic}}
\]
\noindent where $O_{ic}$ and $E_{ic}$ are the observed and the expected number of times the app $i$ was used in context $c$.
To compute $E_{ic}$ we first compute the fraction of times all the apps were used in context $c$ and multiply this fraction by 
the total number of times the app $i$ was used. We use the number of distinct values in the contextual dimension $C \ni c$ as the degree of freedom and compute the p value of the chi square
statistic. We use up to 3 real-time contextual factors of the user with the highest $p$ value and only if $p<0.1$ and $O_{ic} - E_{ic} > 0$.

\section{Studying Frapp\'e In-The-Wild}
In recent years researchers have begun to explore the use of app markets like Google Play as a means of recruiting participants and running large-scale mobile user studies in-the-wild \cite{McMillan:2010,Niels:2011,SahamiShirazi:2013}. While this approach offers a number of benefits, in particular related to the amount of quantitative usage data collected, such studies tend to lack qualitative insights. In consequence, researchers in the mobile HCI community have proposed \emph{hybrid} approaches where mobile applications are evaluated both globally via large-scale app market deployments and concurrently in smaller-scale, local studies \cite{Morrison:2012}. Such approaches lead to richer insights. We opted to evaluate Frapp\'e using a similar approach which involved: (a) releasing the Frapp\'e app globally via Google Play to attract a large user-base enabling us to gather interesting usage statistics; and (b) concurrently running a smaller-scale study in the UK with 33 participants enabling us to gather insights about participant subjective perceptions and experiences with the recommendations provided. In the following sections we describe each deployment in more detail. 

\begin{figure}[h!]
 \centering
  \includegraphics[width=0.98\linewidth]{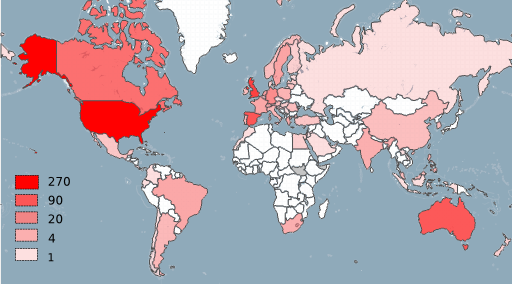}
 \caption{Distribution of downloads around world. \label{fig:map}}
 \end{figure}
 
\subsection{Large-Scale Deployment via Google Play}

Frapp\'e was deployed in the Android Market on the Feb 15th, 2013. We actively advertised the
app among our friends, colleagues and in online social forums such as Reddit. In this paper
we report the usage results derived from the data collected during a 2 month period from Feb 15th - Apr 22nd 2013. 
During this timeframe the app was installed on 1000 mobile devices, equating to 340 different Android user agents.The distribution of user locations is
shown in Figure~\ref{fig:map}. We had users from 37 countries:
41\% of the user base came from the USA, 13\% from UK, 10\% from Spain, 9\% from Australia. Note that majority of users were from English speaking countries. We also observed the primary language of the recorded apps was English (91\% of apps). These are followed by 1.1\% in Chinese , 1.1\% in Spanish and 0.8\% in German.  In total we collected approx. 351K data tuples in the form of $user \times app \times context \times app~usage~count$, where possible values of contexts are listed in Table~\ref{tab:inputs}. 

\subsection{Small-Scale In-Situ Study}
In this section we describe a 3-week field study of Frapp\'e with 33 Android users.

\subsubsection{Participants}
Using an external recruitment company based in the UK, we recruited a diverse group of active Android users.
We define an active Android user as a user who has at least five mobile apps installed on their phone and who uses at least one of these apps once per week.
 Participants were required to own a smartphone running Android OS version 2.2 or higher. 
We recruited 33 participants in total, 21 male and 12 female with varied age ranges: 4 participants were between 18-24, 11 participants were between 25-30, 6 participants were between 31-34, while the remaining 12 participants ranged in age between 35-44. 80\% of participants had 10 or more apps installed, while the remaining 20\% of users had 5-9 apps installed.

\subsubsection{Procedure}
\begin{table*}[t]
\small
\begin{tabular}{lll | lll | lll}
Application & Category & \#\textbf{Installed} & Application & Category & \#\textbf{Viewed} & Application & Category & \#\textbf{Used}\\
\hline
Time Lapse		& Photo	&	25		& Flipboard		& News		& 103	& Chrome		& Browser		& 170K\\
Bubble Shooter 	& Game	&	11		& Expedia Hotels	& Travel		& 49		& Gmail		& Email		& 145K\\
Pocket			& News	&	11		& Time Lapse		& Photo		& 49		& WhatsApp	& Messenger	& 136K \\
Bridge the Wall		& Game	&	10		& Bubble Shooter	& Game		& 48		& Facebook	& Social		& 123K\\
Clean Master		& Tool	&	9		& Firefox			& Browser		& 44		&GO Launcher EX	& Launcher		& 94K\\
\end{tabular}
\caption{Most installed v.s. viewed v.s. used applications. \label{tab:used}}
\end{table*}

We used a moderated online community forum to gather subjective insights from our participants. This online qualitative research approach offers a number of compelling advantages: First, online communities run over a prolonged period of time which allows in-depth interaction and collaboration, thus enabling researchers to review and probe topics in more detail and refine questions as the study progresses. Given that the duration of participant interaction with the researchers in an online community is typically longer than in face-to-face methods, this approach can lead to more qualitative insights and richer information. Second, online communities are more cost effective than face-to-face discussions.

Participants were asked to install the Frapp\'e application on their mobile phone and to use it for a period of 3-weeks. During that time participants accessed a closed, online community forum each evening to answer specific questions about their experiences with Frapp\'e. A new topic was posted to the forum 6 out of 7 days per week (\emph{i.e.} everyday except Sundays) by a moderator over the study period. Each topic comprised of a set of sub-questions or tasks. Topics included: 

\begin{enumerate}
\item The installation, usability, functionality, ease of use and look and feel of the application; 
\setlength{\parskip}{0pt}
\item The perceived value and/or drawbacks of the app both initially and after prolonged use; 
\setlength{\parskip}{0pt}
\item The quality, relevance and suitability of the recommendations provided; 
\setlength{\parskip}{0pt}
\item Their understanding of context, their experiences and perceptions in receiving context-aware recommendations and their understanding of the explanations that Frapp\'e provided to them; 
\setlength{\parskip}{0pt}
\item Their use of Frapp\'e in different places and situations and the relevance of recommendations in various contextual settings; and finally
\setlength{\parskip}{0pt}
\item Their attitudes towards control, preference settings, privacy and security.
\end{enumerate}

Participant responses were moderated and if more detail was required or responses were unclear, the moderator could probe participants in more detail. 
Participants were able to see the responses of fellow participants and were also free to engage in group-based discussions should they desire. 
At the end of each day, all forum responses were analyzed and underlying themes were identified. In this way findings from each day could feed into the questioning for the following day. This iterative approach was very beneficial and enabled us to gather a rich set of subjective insights (see Results section). 
\section{Results}
\label{sec:results}
In this section we report the analysis of both the large-scale deployment and the small-scale user study.
In total, Frapp\'e was installed by approx. 1000 users, \emph{i.e.} on 1000 different mobile devices. 
During deployment we recorded usage of approx. 24K apps which equates to approx. 15.8 unique apps per user. 
This highlights the large variety of apps available and the uniqueness of the user profile 
but also gives a sense of the magnitude of the app discovery problem.
We cleaned the data and removed users who had problems in retrieving recommendations, or did not send any data to the server due 
to unknown technical difficulties. In the rest of the paper we will report the results based on the usage data from 986 users.

\subsection{Characterizing context-dependent mobile app usage}
A number of researchers have explored context-dependent mobile use in the past. For example, in 2006 Verkasalo conducted a study which investigated differences in mobile service usage in general across a variety of contexts like home, work, etc.  \cite{Verkasalo:2009}. More recently B\"{o}hmer et al \cite{Bohmer11-angry} studied temporal patterns of mobile app usage using a dataset collected from over 4000 Android users between Aug 2010 and Jan 2011. The key focus of that work was on durations, categories and sequences of app usage.
Given that Frapp\'e tracks all apps used by all Frapp\'e users (\emph{i.e.} outside the scope of Frapp\'e), the log data recorded during the large-scale deployment provides us with a rich and varied dataset with which we can investigate 
context-dependent mobile app usage behaviors based on a 2-month snapshot of 2013. Because of the high speed at which the mobile space evolves, 
it's likely that new app usage patterns have emerged since prior related work. In this section we report on this characterization.

In total we identified 2.3M app usage events.
App usage is assumed if the users screen is on and there is an app running in the foreground.
Because of the context logging, we were also able to investigate where people use their apps and in what conditions. We found that 
13.6\% of app usage occurred at home, 3\% at work and 83\% at other locations. 
We infer the home and work location
using a time-based heuristic, \emph{i.e.} the most repeated location where the user is from 1am to 6am is considered to be the home 
and most repeated location between 9am and 6pm is considered to be work. 
We tested this approach with 10 pilot users and obtained an accuracy of 95\%. with only 1 false recognition of a work location, 
(all home locations were all correctly recognized), giving us confidence in the accuracy of the used heuristic. 
Our statistical analysis suggests that the location-based contexts that Frapp\'e identified 
(home, work, other) were of very different nature and are not random ($p<10^{-6}$)\footnote{Note that the inference of locations only takes place after Frapp\'e has access to sufficient usage data, which was 3 days on average}.

\begin{figure}[t]
 \centering
  \includegraphics[width=0.8\linewidth]{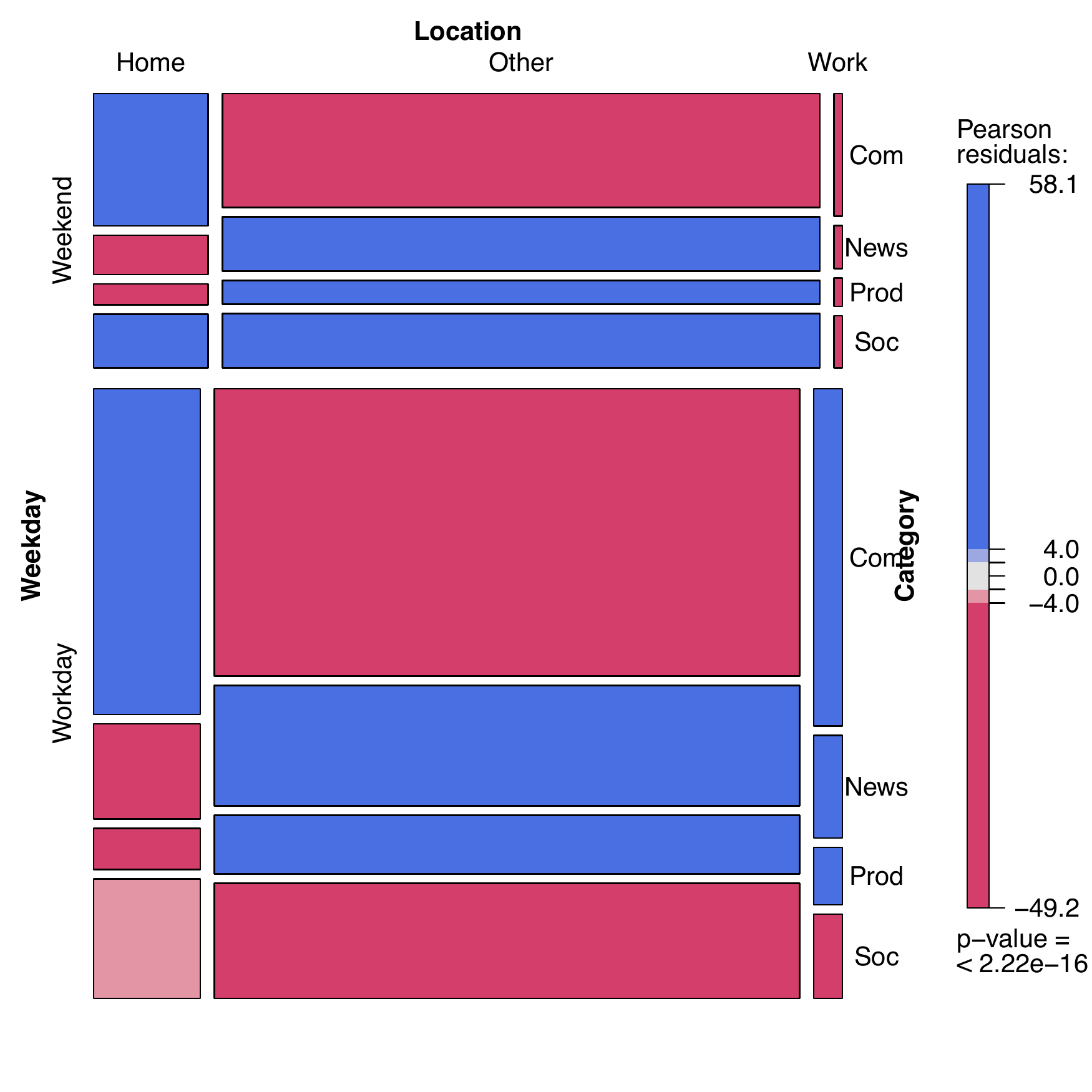}
 \caption{Mosaic plot of the contextual influence on the usage of applications. Usage counts of 4 app categories are displayed (Communication, News, Social, Productivity)  \label{fig:mosaic}}
 \end{figure}

The mosaic plot \cite{friendly94-mosaic} in Figure~\ref{fig:mosaic} helps to visualize the influence of context on the usage of apps for 4 popular app categories 
(Social, Productivity, News and Communication). 
We show location context (home, work, other),
the 4 app categories and temporal context in the form of weekend vs workday. 
The width of each tile is proportional to the marginal frequency of the location dimension, given time.
If there is no dependency on time, the width of the corresponding tiles (\emph{e.g.}, work$|$weekend vs work$|$workday) in these two contexts would be the same.
The red color indicates significant values (using $\chi^2$ statistic) \emph{below} the average, and blue, significant values \emph{above} the average.
Hence, all the cells except for the bottom left cell 
--- Social apps used at home on a workday ---
denote highly statistically significant results. 

The Figure provides rich insights regarding the contextual usage of apps with respect to location, time and category, showing that app usage is highly context dependent (at least for these popular 4 app categories).
Communication apps are by far the most frequently used app category both on weekends and weekdays and particularly while at home. 
Apps across all 4 categories are used less than average at work on the weekend, which seems intuitive. 
Social and Communication apps are used significantly more than average while at home on the weekend, whereas 
Productivity and News apps are used significantly more during workdays both at work and in other locations. 
Interestingly, on workdays, at home Communication apps are more frequently used, when compared to Social, Productivity or News apps. 

\begin{figure*}
  \begin{center}
    \subfigure[Chrome]{\label{fig:}
      \includegraphics[width=0.3\linewidth]{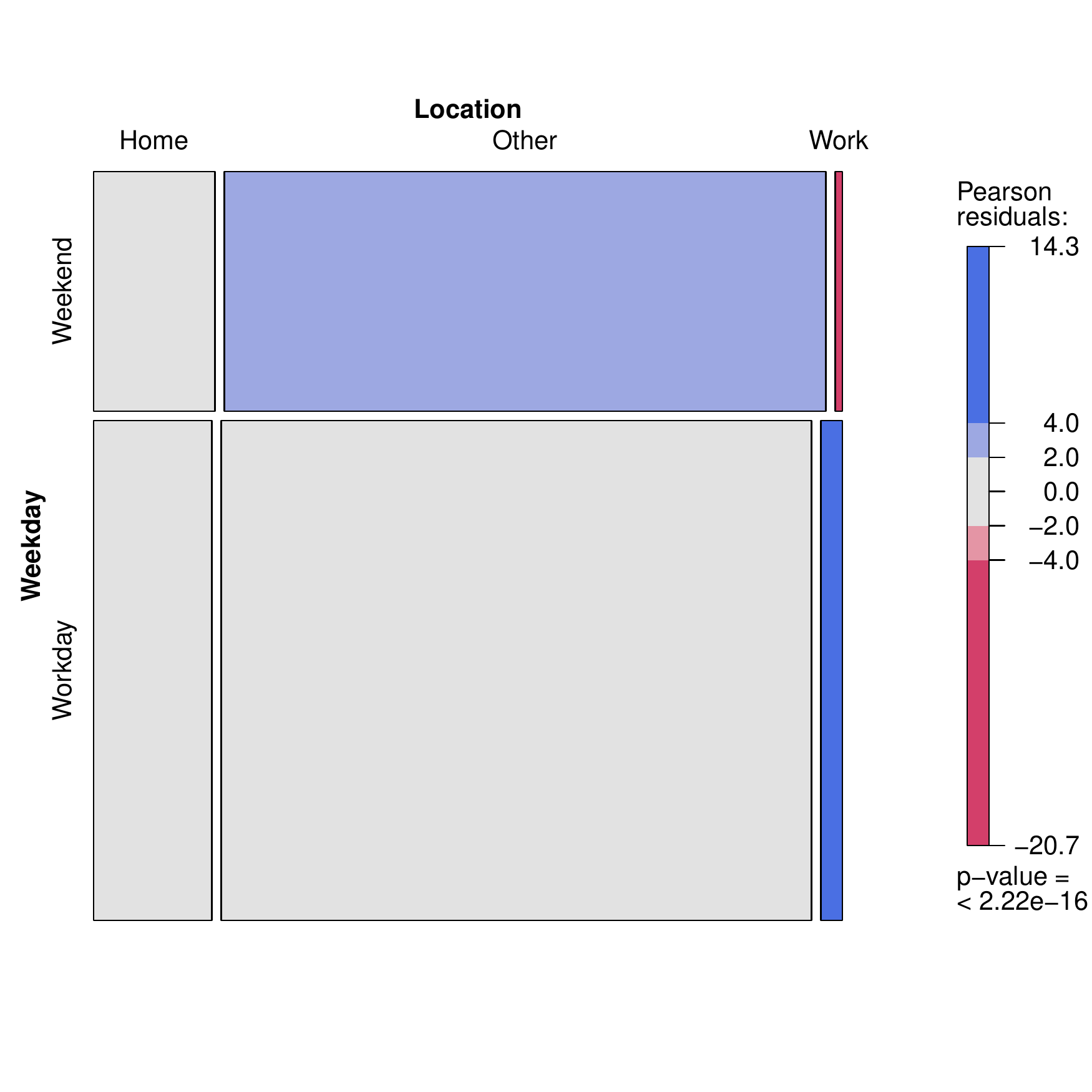}}
    \subfigure[Facebook]{\label{fig:}
      \includegraphics[width=0.3\linewidth]{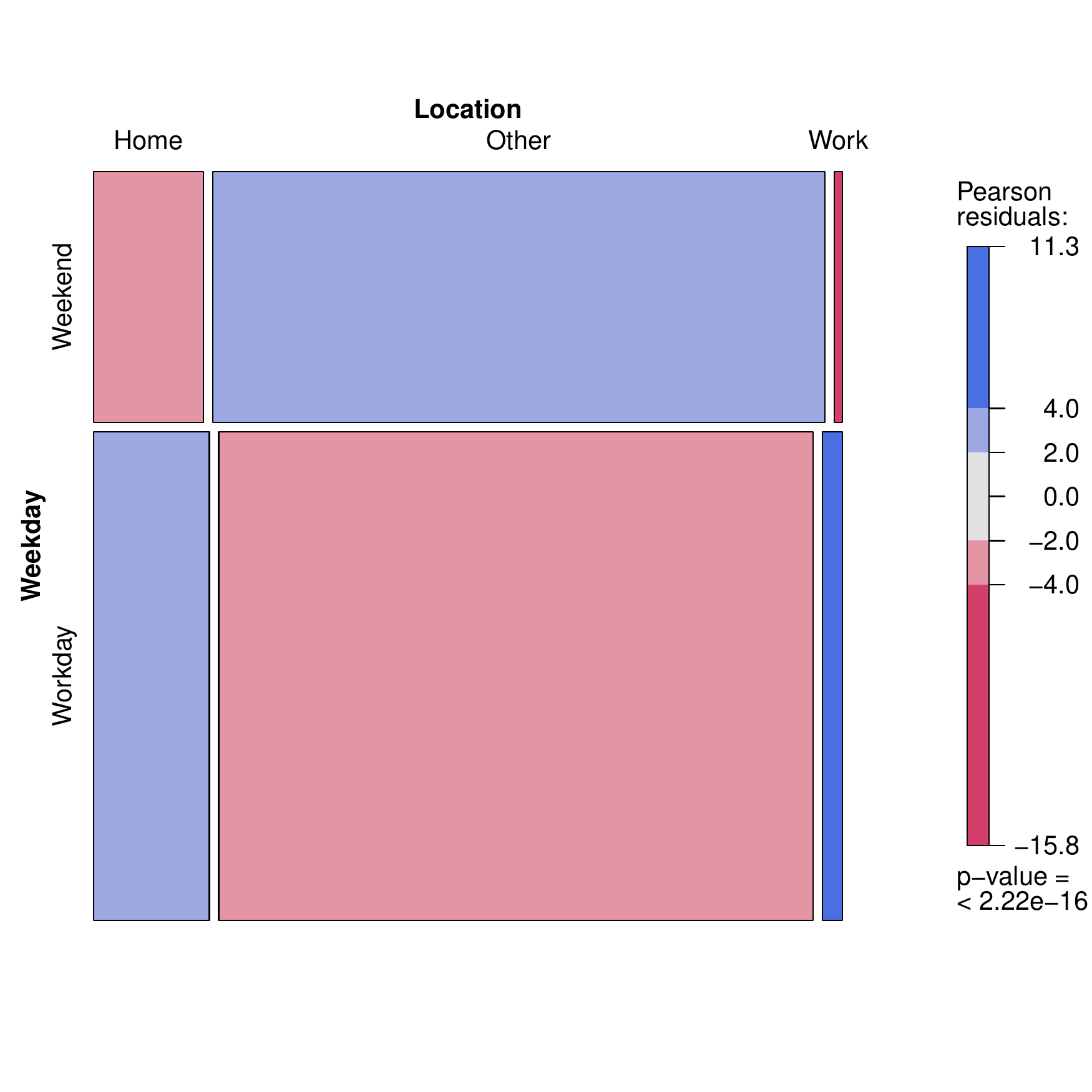}}
    \subfigure[Whatsapp]{\label{fig:}
      \includegraphics[width=0.3\linewidth]{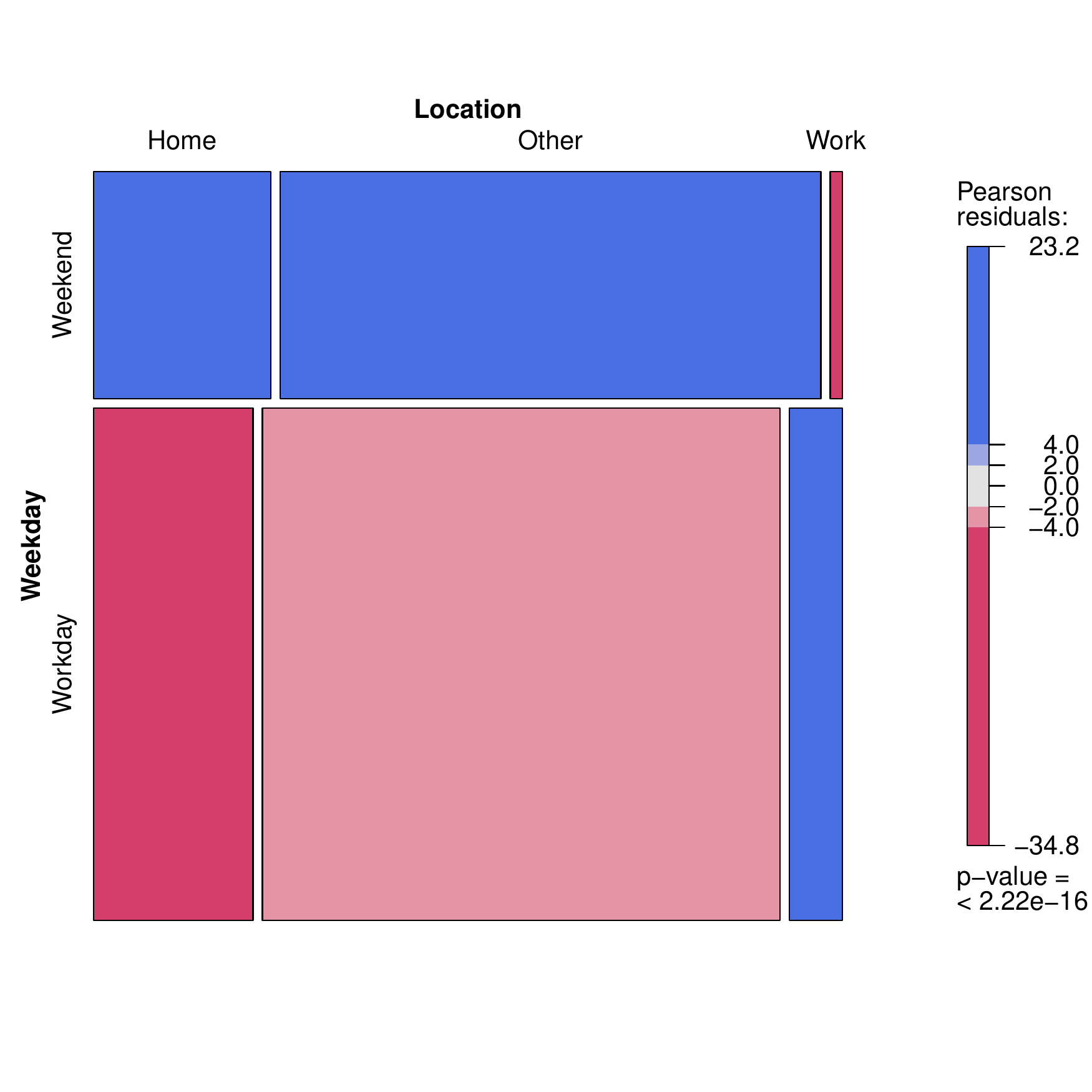}}
     \end{center}
  \caption{App usage in various locations and day of the week.}
  \label{fig:mosaic-apps}
\end{figure*}

While our results highlight contextual influence in terms of categories of apps, we also analyzed the usage of 
three individual popular mobile apps: Chrome (indicative of Web browsing, online information access), 
Facebook (Social app) and Whatsapp (Communication, one of the most popular mobile instant messenger apps). 
Figure~\ref{fig:mosaic-apps} shows that usage of Chrome is the least dependent on contextual factors, whereas, usage of Whatsapp is highly contextual with
high statistical significance. The most interesting observation is the contrasting usage patterns between Whatsapp and Facebook at home during a working day.
Facebook is used more than average on working days while at home, whereas Whatsapp is used significantly less than average in that context. Conversely, 
Facebook is used less than average on weekends while at home and Whatsapp is used significantly more than average in that context.

\subsection{Insights from the Large-Scale Deployment}

This section focuses on findings from the large-scale deployment of Frapp\'e.
In total we logged 4437 sessions of Frapp\'e usage. Each user had logged from 1 to 218 sessions (avg: 4.5, std: 13).
In order to help us assess the effectiveness of the recommendations provided by Frapp\'e, we adopt some of the approaches on usage-centric evaluation app recommendations proposed in \cite{DBLP:conf/iui/BohmerGK13} and compute app \emph{conversion rates} at two stages of app engagement. 
The first is viewed-to-installed (\emph{i.e.} where the user views the app detail page of the recommended app and opts to install the app); the second is installed-to-direct usage.
In total we had 3726 \emph{view app detail} events and 714 of these led to an install of an app. This equates to a 19\% conversion rate and is similar to 
the results reported in \cite{DBLP:conf/iui/BohmerGK13} by a context-aware method. The \emph{view-to-installation} conversion rate 
by a popularity method reported in \cite{DBLP:conf/iui/BohmerGK13} is only 9\%.
Hence, the context-awareness provided by Frapp\'e's recommendations is better than
a standard app popularity method. To investigate further, we explored the conversion rates by app category 
(for categories that had at least 10 installs, to have more reliable statistics).
Games (31\%) have highest conversion rates, followed by news (26\%) and music (23\%). The lowest conversion rates
were for Communication (11\%) and Social (16\%). We believe that games are a much easier category to recommend than social apps as games are more 
fad-like in nature and mobile users are likely to be more comfortable with trying different new games rather than switching to a new communications tool.

Next, we investigated \emph{short term engagement}, that is, if users used the app \emph{directly after the installation}. We obtained a 57.8\% conversion rate compared to 30\%-50\% reported
in \cite{DBLP:conf/iui/BohmerGK13}. 
We further analyzed the logs to get better insights on the success of the recommendations.
On average, users installed 0.16 applications per Frapp\'e session, \emph{i.e.}, they installed 1 recommended app every 6.25 sessions of using Frapp\'e.

In almost every session, users looked at the details of a recommended app (on average 0.84 recommended app views per session). 
Users installed from 0 to 17 recommended apps
(avg: 0.7, std: 1.82). The top installed and viewed \emph{applications} can be seen in Table~\ref{tab:used} together with their category.
We also aggregated the statistics of the most used and viewed apps to determine the most installed and used \emph{categories}.
The most installed categories were \emph{Tools}, \emph{Social} and \emph{Communication} and the most viewed categories were \emph{Communication}, \emph{Social} and \emph{Travel}. 
This is a surprising result, given that Social and Communication apps had the lowest conversion rates.
\subsection{Small-Scale Study}
Due to technical difficulties associated with acquiring the participants device ID, we could not retrieve the log data from 7 participants from the small-scale study.
While we have \emph{qualitative} insights for 33 participants, the final number of participants for whom we have mobile usage data and could do \emph{quantitative} analysis is 26.
In total we collected 646 sessions (min:2,  max:85 sessions, avg:25.5, std: 18.53 per user) from these 26 participants. They installed between 0 and 16 recommended apps (avg: 2.56, std: 3.9). On average users installed 0.1 applications per session (compared to 0.16 in the large deployment) and viewed the details of 0.68 recommended apps per session.
We looked into the conversion rates for this population. The users had 14.4\% view to install conversion and 34.7\% direct usage conversion rates. This is lower than the statistics of the large scale
deployment.

Over the 3-week study period, the forum resulted in $> 2,300$ posts in the form of participant answers to questions. Note that all examples provided in this paper are actual participant responses from the forum. The vast majority of participants found Frapp\'e very easy to use: 
32 participants rated the \emph{ease of use} of the app as 4 or 5 on a 5-point scale, with only one participant providing a neutral rating of 3. The simple, clean layout and intuitive navigation made Frapp\'e appear accessible and user-friendly. Participants were very positive about the concept behind Frapp\'e and believed that it could genuinely save them time, 
\emph{e.g.}: \emph{\sffamily ``I really like it. Its got so many different features to keep me entertained its rather fun looking through the different recommendations."} and \emph{\sffamily ``I like that it explains why its recommending each app (time of day, similar apps), I like that the apps change"} 
 \emph{\sffamily ``I think its great, saves time plus recommends other apps you may like. I like it, good idea"}
Participants overwhelmingly reported that Frapp\'e recommended apps that they had not heard of, or were unlikely to find otherwise, \emph{e.g.}, \emph{\sffamily ``I would say that around 80 per cent of the apps I have been recommended I had never come across before"} and \emph{\sffamily ``The vast majority of these apps I would not have found or considered if they had not been recommend"}.

However and despite installing some of the recommended apps provided by Frapp\'e, in general, participants were quite discerning and selective when installing apps. 
Reasons for not installing recommended apps included: (1) the recommendations did not take their fancy at the specific time, 
(2) cynicism around new apps and a desire to review them first, 
(3) not necessarily perceiving a real benefit and 
(4) limited phone memory.  
For users who did install a few recommended apps we found many positive comments related to their relevance, \emph{e.g.} \emph{\sffamily ``....I was just arriving on a train and I was recommended a taxi app which was perfect. I would usually use Google to search for a local taxi company when I'm traveling for work."}, and \emph{\sffamily ``As far as recommended apps at home they did seem quite relevant. My instruments are at home and so the metronome app was kindly welcomed."}  

One of the most important issues expressed by participants was that they could not see how they could improve or directly influence the recommendations provided, 
\emph{e.g.} \emph{\sffamily ``I don't see how I can improve my recommendations. The same stuff is always recommended, but I have no way of telling it that I don't want it"}. 
While Frapp\'e provides end-user recommendations in a automatic manner, many users expressed a desire to control what was recommended, how apps were recommended and to control the situations and categories in which those apps were recommended. For example, \emph{\sffamily ``I would like the opportunity to tell the app about the things I like and dislike, so that it can tailor my recommendations accordingly"}. 

One of our concerns with deploying the Frapp\'e was privacy. Given that context-aware systems use more information than 
classic RS, a lot of this information could rise significant privacy concerns for the user. As is shown in
Table~\ref{tab:inputs} we use location, time and install/usage information for apps. 
We found that many participants did not understand the exacting contextual data points used to generate their recommendations. In addition we found that the consensus reached by our participants was that if the recommendations improve and provide value, they do not have any problems sharing their location information. However, there needs to be clear benefit for them to share such information. 

\section{Lessons Learned}
\label{sec:lessons}
In this section we outline the key learning outcomes from our field studies of Frapp\'e. 
Some of these findings corroborate previous work, while others complement it. We believe that these lessons would be beneficial to the MobileHCI community, particularly to researchers and practitioners developing mobile systems that include recommendations. 

\subsection{Explaining the large with the small}
As suggested by Morrison et al. \cite{Morrison:2012} in hybrid approaches to mass participation trials we can use the small-scale study to explain the large-scale deployment.  According to the \emph{usage-centric} evaluation framework proposed by  \cite{DBLP:conf/iui/BohmerGK13}, the Frapp\'e system performs well.We find higher app conversations rates (19\%) in terms of \emph{views to installs} when compared to standard popularity based metrics (9\%) and app aware filtering (7\%). We also find higher conversion rates in terms of \emph{installs to direct usage} (57.8\% vs. 30-50\%) when compared to the approaches evaluated in \cite{DBLP:conf/iui/BohmerGK13}. The key issue, however, is that while, Frapp\'e performs well based on this \emph{usage-centric evaluation approach}, the small scale study highlighted some important drawbacks of the system that negatively effected end-user experiences. We will discuss each of this key issues below, however our first key lesson relates to the importance of using a hybrid approach in order to get a more complete picture of user experiences and perceptions of mobile app recommendations. 

\subsubsection{Avoid Highly Irrelevant Recommendations}
Many (21 of 33) participants in the small-scale study 
reported issues related to irrelevant recommendations at least once
during the study. We believe this is mainly due to the lack of
data to train our complex models (cold-start problem). Most of these issues were related to receiving app
recommendations in a foreign language, \emph{e.g.} \emph{\sffamily ``I have now
received 3 different apps in foreign languages. I believe them to be all
German but as I don't speak this language, I cannot say 100\% it is!"}.
Other comments signaled that Frapp\'e lacked a basic
understanding of the target audience for some of the apps, \emph{e.g.}  \emph{\sffamily ``I was offered
the 'My Pregnancy Today' app, because I have a stopwatch app. This isn't
really relevant as I am a male"}.  These highly
irrelevant recommendations have an extremely negative effect of end-user experiences. 
Considering that existing
evaluations of RS focus mainly on precision/recall on relevant recommendations in their evaluations,
we suggest that future evaluations of RS, in particular for mobile systems, take these highly irrelevant items into account. 
Paul Lamere coined a test for such irrelevant items called the \emph{WTF test}\footnote{\url{http://musicmachinery.com/2011/05/14/how-good-is-googles-
instant-mix/}} in which a \emph{WTF score} is computed by summing up
highly irrelevant items in the top-N recommendation
list. 
Thus we propose to adopt RS evaluation metrics that take into account 
the \emph{severity} of the mistakes (via WTF scores or other approaches), not just the number of mistakes.

\subsubsection{Learn Quickly}
Also of critical note is the time-frame a RS has to
prove itself as effective. It is well known that new users first assess 
if they can trust a system and tend to be quite
critical when doing so \cite{Ribeiro12-pareto,Herlocker04-evaluating}. 
In our small-scale study the participants came to a consensus that they would
give an app recommender between 1 week and 1 month to prove itself.
However, when we analyzed the usage log
data from the large-scale deployment we noticed a very different pattern. 
Specifically, unsatisfied users did not wait for a month, but uninstalled apps within an hour from its installation. 
While we don't know entirely why app uninstalls took place, it is reasonable to assume that uninstall actions are 
a signal of disinterest or dislike of the recommended app. Thus it seems that mobile app RS have very little time to prove their worth, which means that it is
imperative to get the recommendations right immediately. Moreover, we would argue that short-term uninstall conversion rate should also 
be a key part of any usage-centric app recommendation evaluation measures \cite{DBLP:conf/iui/BohmerGK13}.

\subsubsection{Automatic Discovery vs Explicit Control.}

Our aim with Frapp\'e was to build a fully automated app discovery system, \emph{i.e.}, where recommendations
are provided by observing implicit user actions without the need for explicit input or feedback from the user.
Participants of the small-scale study felt that the dialogue with Frapp\'e was largely one-way, with minimal capacity for them to control their experience. 
When participants were asked about what features they would like to include in Frapp\'e, the majority of these were related to controlling the recommendations and their preferences, e.g. 28 of the 33 participants indicated that would like a \emph{blacklist} facility, \emph{i.e.} the ability to tell Frapp\'e when they do not like a recommended app and ensure that the 
app is never recommended again.
19 of the 33 participants indicated a need for a \emph{favorites list} where they can add apps they are interested in for later installation. 
In addition participants wanted to personalize the service by telling Frapp\'e more details about them, \emph{e.g.},  \emph{\sffamily{``It would have been handy to tell the app your workdays for example"}} 
Thus, participants wanted a two-way dialogue involving varying levels of feedback from them coupled with the ability to set preferences and control their experiences more explicitly. Providing these capabilities and learning from these preferences would lead to enriched end-user experiences. Critiquing-based RS \cite{DBLP:reference/rsh/McGintyR11} have been developed to precisely address this challenge. 
In future work, we would like to incorporate elements of mixed-initiative systems into Frapp\'e. 

\subsubsection{Use \& Perceptions of Context}
While the majority of users
expected their recommendations to adapt to their current situation, at times Frapp\'e did not meet
their expectations in this regard. 
Users reported:
\begin{enumerate} 
\item Perceiving no adaptation based on their current situation, \emph{e.g.}  \emph{\sffamily ``I have used the app either at home or work and have not noticed 
any difference to be honest."}
\setlength{\parskip}{0pt}
\item Encountering irrelevant recommendations given to their 
current location, \emph{e.g.}  \emph{\sffamily ``When out and about I didn't notice that I was being recommended any travel apps or things related to location at all"}.
\setlength{\parskip}{0pt}
\item Encountering irrelevant recommendations based on temporal context, \emph{e.g.} \emph{\sffamily ``I get apps that are recommended because it is the weekend, I assume that the app thinks everybody is off work at the weekend, for me this isn't the case, my days off change every week."}.
\setlength{\parskip}{0pt}
\end{enumerate}

Likewise, Frapp\'e attemped to convey its context-awareness in the form of context-sensitive explanations (See Figure~\ref{fig:ss8}). 
It used a broad definition of why the recommendation was provided, 
by saying ``Recommended because your current situation is: Afternoon, Barcelona, Spain''. 
These explanations successfully communicated the use of the context to the end-user. However, some users (18\%) perceived these explanations as
too generic. \emph{\sffamily ``Android Music player... It said that the day of the week is Wednesday, the country is UK, and it is a workday\ldots on those reasons it could have recommended anything!"}.
While contextual data can help the algorithm to build a better predictive model, such data did not always help in explaining the recommendations.

These findings reveal that users understand the concept of context dependent adaptations and already have expectations about how a system should adapt based on their context. 
Moreover, the patterns of app usage in various contexts show that context-awareness is essential while modeling mobile users and their behaviors.
However, user expectations that a contextual service will adapt to their current context are very high and 
are likely to be different for each user, therefore, they can be difficult to fulfill.

\subsection{Evaluation of CARS in the Mobile Space}

We have provided comparisons to the usage-centric app recommendation evaluation framework proposed by \cite{DBLP:conf/iui/BohmerGK13} 
and have obtained good performance results. However, based on rich user insights collected from the small-scale study and our experiences via the large-scale deployment, we have a suggestion to enrich the framework proposed in \cite{DBLP:conf/iui/BohmerGK13} which we believe will help researchers working in the mobile CARS space to better evaluate their mobile app recommenders.

We believe that the direct usage conversion measure might not be the best measure to assess the \emph{quality of app recommendations}. 
Usage that occurs straight after the installation of an app is 
likely caused by curiosity rather than by the quality of the app recommendations. Mobile users must launch/open an app once installed to see what the app holds within.
We would argue that a more important measure of \emph{quality (failure)} is the \emph{direct uninstall} of the application
that was just installed. A variation of this measure is used by Appaware (personal communication with the appaware team) and Google 
play\footnote{https://developers.google.com/events/io/sessions/326335584} as one of the most important signals for the quality of an application.
\begin{figure}
  \begin{center}
    \subfigure[Days]{\label{fig:uninstall:day}
      \includegraphics[width=0.4\linewidth]{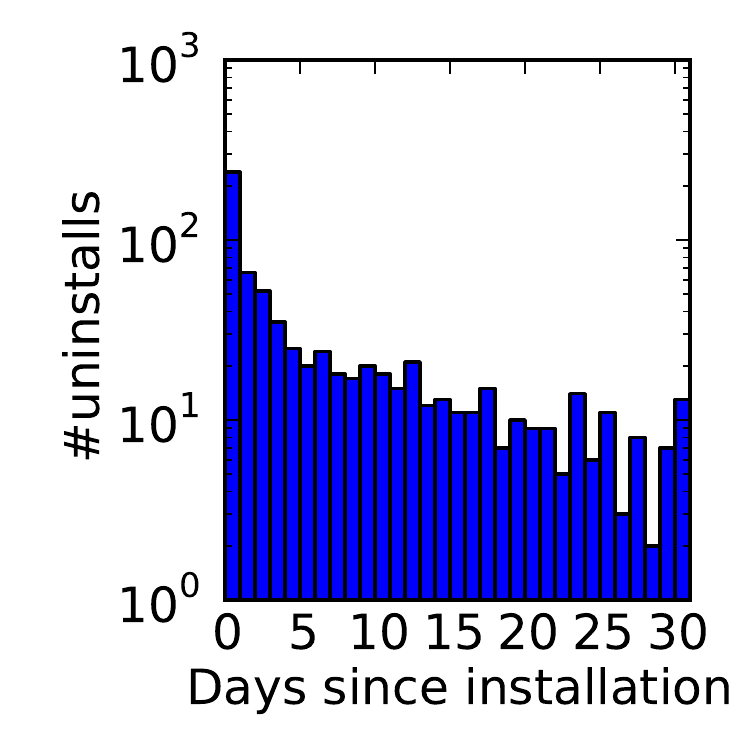}}
    \subfigure[Hours]{\label{fig:uninstall:hour}
      \includegraphics[width=0.4\linewidth]{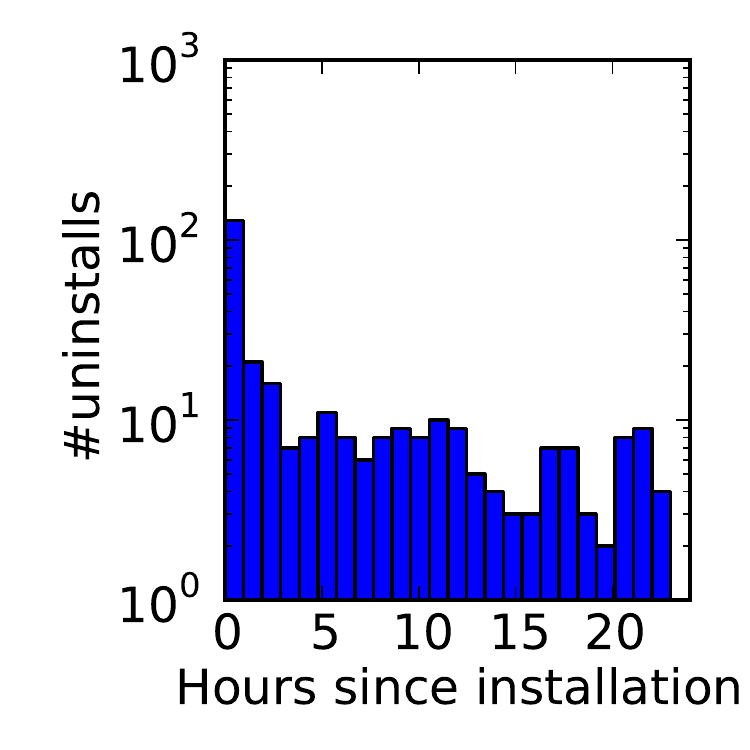}}
   \end{center}
  \caption{Time since uninstall of the Frapp\'e app.}
  \label{fig:uninstall}
\end{figure}

To showcase this measure, we computed the uninstall rates for the Frapp\'e app.
Figure~\ref{fig:uninstall:day} shows that the majority of
uninstalls of Frapp\'e occurred \emph{during the first day}. Actually, 
most users uninstalled Frapp\'e during the first hour (see Figure~\ref{fig:uninstall:hour}),
others within the first minutes. This highlights the fact that users evaluate the
perceived quality and benefit of the RS (or any other app) within a very
short period of time. As such if you provide the wrong recommendations within the
first minutes, it is likely that the user's loyalty to the RS will be
threatened. By looking at not only these uninstall rates, but also the time between install and uninstall, users can better assess the quality and effectiveness of their app recommendations.

\subsection{Predicting Usage vs. Recommending Apps}
In most RS domains, usage of an item is taken as
a positive signal that the user liked the item in question. This signal
can be noisy, but on average, consumption patterns can be used to
generate reliable recommendations \cite{koren08-implicit}. 
This is a reasonable assumption in many domains where recommendations are 
commonplace (\emph{e.g.} movies, music, etc...). However,
it might not be appropriate in the app recommendation domain, or at least 
not for all categories of apps. 

In fact, directly
employing app usage data can lead to poor recommendations because usage patterns and app installation patterns 
can be very different. Apps are typically installed to satisfy certain needs, which are highly dependent on the context. 
An app (\emph{i.e.} Compass) might be used rarely, but might be of great importance to the user.
Table~\ref{tab:used} shows the top installed, viewed and used apps.
Most of the top apps are browsers, launchers\footnote{A launcher presents the main view of the phone and is responsible for starting other apps and hosting live widgets.} and
communication tools. Given that most users already have their favorite of these tools installed on their mobile phones, recommending similar apps, 
even if these apps are not known to the user, might not be the most appropriate action to take. 
Even though recent work has shown that users install and use more than one app within the same category 
\cite{SahamiShirazi:2013}, in our experience, the participants of the small-scale study did not 
appreciate being recommended apps within the same category of apps that they frequently used.

One might argue that installation data rather than usage data could be used to train the RS models. 
However, after our 2 month deployment, we only recorded 714 install
actions, compared to 2.3M apps usage actions. 
Thus it is unlikely that a service like Frapp\'e will ever record a sufficient number of install actions to use such data for training the RS models.  
As such we believe there is a great opportunity for future research on novel
implicit algorithms to provide enhanced mobile app recommendations.


\section{Conclusions}
We have presented insights and lessons learned from 
 a hybrid study of a context-aware
app recommendater called Frapp\'e. 
Our results indicate that contextual variables, such as location and time are very
important signals for modeling app usage and providing recommendations.
However, feedback collected in the small scale user study, 
shows that while users understand the value of 
context dependent adaptation, their expectations in this regard are also very high.
We provide a set of lessons learned which outline important considerations in designing, deploying and evaluating mobile context-aware RS in-the-wild with real users.


\bibliographystyle{abbrv}

{\footnotesize
\bibliography{recsys2013}}

\begin{thebibliography}{10}

\bibitem{Adomavicius10-handbook}
G.~Adomavicius and A.~Tuzhilin.
\newblock Context-aware recommender systems.
\newblock In F.~Ricci, L.~Rokach, B.~Shapira, and P.~B. Kantor, editors, {\em
  Recommender Systems Handbook}, pages 217 -- 250. Springer, 2011.

\bibitem{Baltrunas11-journal}
L.~Baltrunas, B.~Ludwig, S.~Peer, and F.~Ricci.
\newblock Context relevance assessment and exploitation in mobile recommender
  systems.
\newblock {\em Personal Ubiquitous Comput.}, 16(5):507--526, June 2012.

\bibitem{DBLP:conf/iui/BohmerGK13}
M.~B{\"o}hmer, L.~Ganev, and A.~Kr{\"u}ger.
\newblock Appfunnel: a framework for usage-centric evaluation of recommender
  systems that suggest mobile applications.
\newblock In J.~Kim, J.~Nichols, and P.~A. Szekely, editors, {\em IUI}, pages
  267--276. ACM, 2013.

\bibitem{Bohmer11-angry}
M.~B{\"o}hmer, B.~Hecht, J.~Sch{\"o}ning, A.~Kr{\"u}ger, and G.~Bauer.
\newblock Falling asleep with angry birds, facebook and kindle: a large scale
  study on mobile application usage.
\newblock In {\em Proceedings of Mobile HCI}, pages 47--56, 2011.

\bibitem{Bohmer:2013mobil}
M.~B\"{o}hmer and A.~Kr\"{u}ger.
\newblock A study on icon arrangement by smartphone users.
\newblock In {\em Proceedings of the SIGCHI Conference on Human Factors in
  Computing Systems}, CHI '13, pages 2137--2146, New York, NY, USA, 2013. ACM.

\bibitem{Church-2009}
K.~Church and B.~Smyth.
\newblock Understanding the intent behind mobile information needs.
\newblock In {\em Proceedings of IUI '09}, pages 247--256. ACM, 2009.

\bibitem{Cremonesi-2011-eval}
P.~Cremonesi, F.~Garzotto, S.~Negro, A.~Papadopoulos, and R.~Turrin.
\newblock Looking for good recommendations: A comparative evaluation of
  recommender systems.
\newblock In P.~Campos, N.~Graham, J.~Jorge, N.~Nunes, P.~Palanque, and
  M.~Winckler, editors, {\em INTERACT 2011}, volume 6948 of {\em Lecture Notes
  in Computer Science}, pages 152--168. Springer Berlin Heidelberg, 2011.

\bibitem{Davidsson:2011}
C.~Davidsson and S.~Moritz.
\newblock Utilizing implicit feedback and context to recommend mobile
  applications from first use.
\newblock In {\em Proceedings of the 2011 Workshop on Context-awareness in
  Retrieval and Recommendation}, CaRR '11, pages 19--22, New York, NY, USA,
  2011. ACM.

\bibitem{friendly94-mosaic}
M.~Friendly.
\newblock Mosaic displays for multi-way contingency tables.
\newblock {\em JASA}, pages 190--200, 1994.

\bibitem{Girardello:2010}
A.~Girardello and F.~Michahelles.
\newblock Appaware: which mobile applications are hot?
\newblock In {\em Proceedings of MobileHCI'10}, pages 431--434. ACM, 2010.

\bibitem{Niels:2011}
N.~Henze, M.~Pielot, B.~Poppinga, T.~Schinke, and S.~Boll.
\newblock My app is an experiment: Experience from user studies in mobile app
  stores.
\newblock {\em International Journal of Mobile Human Computer Interaction
  (IJMHCI)}, 3(4):71--91, 2011.

\bibitem{Herlocker04-evaluating}
J.~L. Herlocker, J.~A. Konstan, L.~G. Terveen, and J.~T. Riedl.
\newblock Evaluating collaborative filtering recommender systems.
\newblock {\em ACM Transactions on Information Systems}, 22:5--53, 2004.

\bibitem{koren08-implicit}
Y.~Hu, Y.~Koren, and C.~Volinsky.
\newblock Collaborative filtering for implicit feedback datasets.
\newblock In {\em Proceedings of ICDM '08}, pages 263--272. IEEE Computer
  Society, December 2008.

\bibitem{Jannach:2009}
D.~Jannach and K.~Hegelich.
\newblock A case study on the effectiveness of recommendations in the mobile
  internet.
\newblock In {\em Proceedings of the Third ACM Conference on Recommender
  Systems}, RecSys '09, pages 205--208, New York, NY, USA, 2009. ACM.

\bibitem{Karatzoglou12-appwall}
A.~Karatzoglou, L.~Baltrunas, K.~Church, and M.~B{\"o}hmer.
\newblock Climbing the app wall: enabling mobile app discovery through
  context-aware recommendations.
\newblock In {\em Proceedings of CIKM'12}, pages 2527--2530, 2012.

\bibitem{Knijnenburg2012}
B.~P. Knijnenburg, M.~C. Willemsen, Z.~Gantner, H.~Soncu, and C.~Newell.
\newblock Explaining the user experience of recommender systems.
\newblock {\em User Modeling and User-Adapted Interaction}, 22(4-5):441--504,
  2012.

\bibitem{Konstan:2012}
J.~A. Konstan and J.~Riedl.
\newblock Recommender systems: from algorithms to user experience.
\newblock {\em User Modeling and User-Adapted Interaction}, 22(1-2):101--123,
  Apr. 2012.

\bibitem{DBLP:reference/rsh/McGintyR11}
L.~McGinty and J.~Reilly.
\newblock On the evolution of critiquing recommenders.
\newblock In F.~Ricci, L.~Rokach, B.~Shapira, and P.~B. Kantor, editors, {\em
  Recommender Systems Handbook}, pages 419--453. Springer, 2011.

\bibitem{McMillan:2010}
D.~McMillan, A.~Morrison, O.~Brown, M.~Hall, and M.~Chalmers.
\newblock Further into the wild: running worldwide trials of mobile systems.
\newblock In {\em Proceedings of Pervasive'10}, pages 210--227.
  Springer-Verlag, 2010.

\bibitem{McNee:2006}
S.~M. McNee, J.~Riedl, and J.~A. Konstan.
\newblock Being accurate is not enough: how accuracy metrics have hurt
  recommender systems.
\newblock In {\em Proceedings of CHI '06 Extended Abstracts}, pages 1097--1101.
  ACM, 2006.

\bibitem{Morrison:2012}
A.~Morrison, D.~McMillan, S.~Reeves, S.~Sherwood, and M.~Chalmers.
\newblock A hybrid mass participation approach to mobile software trials.
\newblock In {\em Proceedings of the CHI'12}, pages 1311--1320. ACM, 2012.

\bibitem{Pu:2011}
P.~Pu, L.~Chen, and R.~Hu.
\newblock A user-centric evaluation framework for recommender systems.
\newblock In {\em Proceedings of RecSys '11}, pages 157--164. ACM, 2011.

\bibitem{Ribeiro12-pareto}
M.~T. Ribeiro, A.~Lacerda, A.~Veloso, and N.~Ziviani.
\newblock Pareto-efficient hybridization for multi-objective recommender
  systems.
\newblock In P.~Cunningham, N.~J. Hurley, I.~Guy, and S.~S. Anand, editors,
  {\em RecSys}, pages 19--26. ACM, 2012.

\bibitem{SahamiShirazi:2013}
A.~Sahami~Shirazi, N.~Henze, T.~Dingler, K.~Kunze, and A.~Schmidt.
\newblock Upright or sideways?: Analysis of smartphone postures in the wild.
\newblock In {\em Proceedings of the 15th International Conference on
  Human-computer Interaction with Mobile Devices and Services}, MobileHCI '13,
  pages 362--371, New York, NY, USA, 2013. ACM.

\bibitem{Verkasalo:2009}
H.~Verkasalo.
\newblock Contextual patterns in mobile service usage.
\newblock {\em Personal Ubiquitous Comput.}, 13(5):331--342, June 2009.

\bibitem{Woerndl:2007}
W.~Woerndl, C.~Schueller, and R.~Wojtech.
\newblock A hybrid recommender system for context-aware recommendations of
  mobile applications.
\newblock In {\em Proceedings of ICDEW '07}, pages 871--878, Washington, DC,
  USA, 2007. IEEE Computer Society.

\bibitem{Yan-2011}
B.~Yan and G.~Chen.
\newblock Appjoy: personalized mobile application discovery.
\newblock In {\em Proceedings of MobiSys'11}, MobiSys '11, pages 113--126. ACM,
  2011.

\end{thebibliography}

\end{document}